\documentclass[runningheads]{llncs}

\usepackage{graphicx}

\usepackage{amsmath}
\usepackage{amssymb}
\usepackage{xspace}
\usepackage{epstopdf}

\newcommand{\sign}{\boldsymbol{s}}

\newcommand{\img}{\boldsymbol{x}}
\newcommand{\imgd}{{\img'}}

\newcommand{\rfi}{\img_{\mathrm{RF}(i)}}
\newcommand{\rfni}{\tilde{\img}_{\mathrm{RF}({i})}}
\newcommand{\param}{\boldsymbol{\theta}}

\newcommand{\signMMSE}{\sign^{\mathrm{MMSE}}}

\newcommand{\E}[1]{\mathbb{E}\left[ #1 \right]}
\newcommand{\Esub}[2]{\mathbb{E}_{#2} \left[ #1 \right]}

\newcommand{\compactsubsection}[1]{\vspace{0.7mm}\noindent\textbf{#1:}}
\newcommand{\compactsubsubsection}[1]{\vspace{0.4mm}\noindent\textBF{\textit{#1:}}}

\DeclareMathOperator*{\argmin}{argmin}

\newcommand{\etal}{\textit{et al.}\xspace}
\newcommand{\ie}{\textit{i.e.}\xspace}
\newcommand{\eg}{\textit{e.g.}\xspace}

\renewcommand{\tabcolsep}{1.5mm}

\usepackage{graphicx}
\newsavebox\CBox
\def\textBF#1{\sbox\CBox{#1}\resizebox{\wd\CBox}{\ht\CBox}{\textbf{#1}}}

\usepackage{perpage} 
\MakePerPage{footnote} 

\begin{document}
\title{Probabilistic Noise2Void:\\ Unsupervised Content-Aware Denoising
}
%
%
\author{Alexander Krull$^1$, Tom\'{a}\v{s} Vi\v{c}ar$^2$, Florian Jug$^1$ }
\authorrunning{Alexander Krull, Tom\'{a}\v{s} Vi\v{c}ar, Florian Jug}
%
\institute{MPI-CBG/PKS (CSBD)  \and
Brno University of Technology}
\maketitle              
\begin{abstract}
\fussy
Today, Convolutional Neural Networks (CNNs) are the leading method for image denoising.
They are traditionally trained on pairs of images, which are often hard to obtain for practical applications.
This motivates self-supervised training methods such as Noise2Void~(N2V) that operate on single noisy images.
Self-supervised methods are, unfortunately, not competitive with models trained on image pairs.
\sloppy
Here, we present \textit{Probabilistic Noise2Void} (PN2V), a method to train CNNs to predict per-pixel intensity distributions.
Combining these with a suitable description of the noise, we obtain a complete probabilistic model for the noisy observations and true signal in every pixel.
We evaluate PN2V on publicly available microscopy datasets, under a broad range of noise regimes, and achieve competitive results with respect to supervised state-of-the-art methods.

\end{abstract}
%
\section{Introduction}
\label{sec:introduction}
Image restoration is the problem of reconstructing an image from a corrupted version of itself.
Recent work shows how CNNs can be used to build powerful content-aware image restoration (CARE) pipelines~\cite{Weigert2017,Weigert2018usingEtAl,zhang2017beyond,zhang2018poisson,noise2noise,krull2019noise2voidCVPR,batson2019noise2self,Laine2019self-supervised}.
However, for supervised CARE models, such as~\cite{Weigert2018usingEtAl}, pairs of clean and noisy images are required.

For many application areas, it is impractical or impossible to acquire clean ground-truth images~\cite{buchholz2018cryo}.
In such cases, Noise2Noise (N2N) training~\cite{noise2noise} relaxes the problem, only requiring two noisy instances of the same data.
Unfortunately, even the acquisition of two noisy realizations of the same image content is often difficult~\cite{buchholz2018cryo}.
Self-supervised training methods, such as Noise2Void (N2V)~\cite{krull2019noise2voidCVPR}, are a promising alternative, as they operate exclusively on single noisy images~\cite{krull2019noise2voidCVPR,batson2019noise2self,Laine2019self-supervised}. This is enabled by excluding/masking the center (blind-spot) of
the network's receptive fields.
Self-supervised training assumes that the noise is pixel-wise independent and that the true intensity of a pixel can be predicted from local image context, excluding before-mentioned blind-spots~\cite{krull2019noise2voidCVPR}.
For many applications, especially in the context of microscopy images, the first assumption is fulfilled, but the second assumption offers room for improvements~\cite{Laine2019self-supervised}.

Hence, self-supervised models can often not compete with supervised training~\cite{krull2019noise2voidCVPR}.
In concurrent work, by Laine~\etal~\cite{Laine2019self-supervised}, this problem was elegantly addressed by assuming a Gaussian noise model and predicting Gaussian intensity distributions per pixel.
The authors also showed that the same approach an be applied to other noise distributions, which can be approximated as Gaussian, or can be described analytically.

Here, we introduce a new training approach called \textit{Probabilistic Noise2Void (PN2V)}.
Similar to~\cite{Laine2019self-supervised}, PN2V proposes a way to leverage information of the network's blind-spots.
However, PN2V is not restricted to Gaussian noise models or Gaussian intensity predictions.
More precisely, to compute the posterior distribution of a pixel, we combine
$(i)$~a general noise model that can be represented as a histogram (observation likelihood), and
$(ii)$~a distribution of possible true pixel intensities (prior), represented by a set of predicted samples.

Having this complete probabilistic model for each pixel, we are now free to chose which statistical estimator to employ.
In this work we use MMSE estimates for our final predictions and show that MMSE-PN2V consistently outperformes other self-supervised methods and, in many cases, leads to results that are competitive even with supervised state-of-the-art CARE networks (see below).

\section{Background}
\label{sec:background}
\compactsubsection{Image Formation and the Denoising Task}
\label{sec:formation}
An image $\img=(\img_1,\dots,\img_n)$ is the corrupted version of a clean image (signal) $\sign=(\sign_1,\dots,\sign_n)$.
Our goal is to recover the original signal from $\img$, thus implementing a function
$
f(\img) = \hat{\sign} \approx \sign
$.

In this paper, we assume that each observed pixel value $\img_i$ is independently drawn from the conditional distribution $p(\img_i|\sign_i)$ such that
\begin{equation}
    p(\img|\sign)= \prod_{i=1}^n p(\img_i|\sign_i). \label{eq:corruption}
\end{equation}
We will refer to $p(\img_i|\sign_i)$ as \emph{observation likelihood}. It is described by an arbitrary noise model.

\compactsubsection{Traditional Training and Noise2Noise}
\label{sec:traditional}
The function $f(\img)$ can be implemented by a Fully Convolutional Network (FCN) \cite{long2015fully} (see \eg ~\cite{Weigert2017,Weigert2018usingEtAl,zhang2017beyond,noise2noise}), a type of CNN that takes an image as input and produces an entire (in this case denoised) image as output.
However, in this setup every predicted output pixel $\hat{\sign}_i$ depends only on a limited receptive field $\rfi$, \ie a patch of input pixels surrounding it.
FCN based image denoising in fact implements $f(\img)$ by producing independent predictions
$
    \hat{\sign}_i = g(\rfi;\param)  \approx \sign_i
$
for each pixel $i$, depending only on $\rfi$ instead of on the entire image.
The prediction is parametrized by the weights $\param$ of the network.

In traditional training, $\param$ are learned from pairs of noisy $\img^j$ and corresponding clean training images $\sign^j$, which provide training examples
$(\rfi^j,\sign_i^j)$ consisting of noisy input patches $\rfi^j$ and their corresponding clean target values $\sign_i^j$.
The parameters $\param$ are traditionally tuned to minimize an empirical risk function such as the average squared distance
\begin{equation}
\argmin_{\param} \sum_{i=1}^n \sum_{j=1}^m (\hat{\sign}_i^j - \sign_i^j )^2
\label{eq:risk}
\end{equation}
over all training images $j$ and pixels $i$.

In Noise2Noise \cite{noise2noise}, Lehtinen \etal show that clean data is in fact not necessary for training and that the same training scheme can be used with noisy data alone.
Noise2Noise uses pairs of corresponding noisy training images $\img^j$ and $\imgd^j$, which are based on the same signal $\sign^j$, but are corrupted independently by noise (see Eq.~\ref{eq:corruption}).
Such pairs can for example be acquired by imaging a static sample twice.
Noise2Noise uses training examples $(\rfi^j,\imgd^j_i)$, with the input patch $\rfi^j$ cropped from the first image $\img^j$ and the noisy target $\imgd^j_i$ extracted from the patch center in the second one $\imgd$.
It is of course impossible for the network to predict the noisy pixel value $\imgd^j_i$ from the independently corrupted input $\rfi^j$.
However, assuming the noise is zero centered, \ie
$
\E{\imgd_i^j}=\sign_i^j, \label{eq:zero_mean}
$
the best achievable prediction is the clean signal $\sign_i^j$ and the network will learn to denoise the images it is presented with.

\compactsubsection{Noise2Void Training}
\label{sec:noise2void}
In Noise2Void, Krull \etal \cite{krull2019noise2voidCVPR} show that training is still possible when not even noisy training pairs are available.
They use single images to extract input and target for their networks.
If this was done naively, the network would simply learn the identity transformation, directly outputting the value at the center of each pixel's receptive field.
Krull \etal address the issue by effectively removing the central pixel from the networks receptive field.
To achieve this, they mask the pixel during training, replacing it with a random value from the vicinity.
Thus, a Noise2Void trained network can be seen as a function
$
    \hat{\sign}_i = \tilde{g}(\rfni;\param)  \approx \sign_i,
$
making a prediction for a single pixel based on the modified patch $\rfni$ that excludes the central pixel.
Such a network can no longer describe the identity, and can be trained from single noisy images.

However, this ability comes at price. The accuracy of the predictions is reduced, as the network has to exclude the central pixel of its receptive field, thus having less information available.

To allow efficient training of a CNN with Noise2Void, Krull \etal simultaneously mask multiple pixels in larger training patches and jointly calculate their gradients.

\begin{figure}[bt]
\begin{center}
	  \includegraphics[width=1\textwidth]{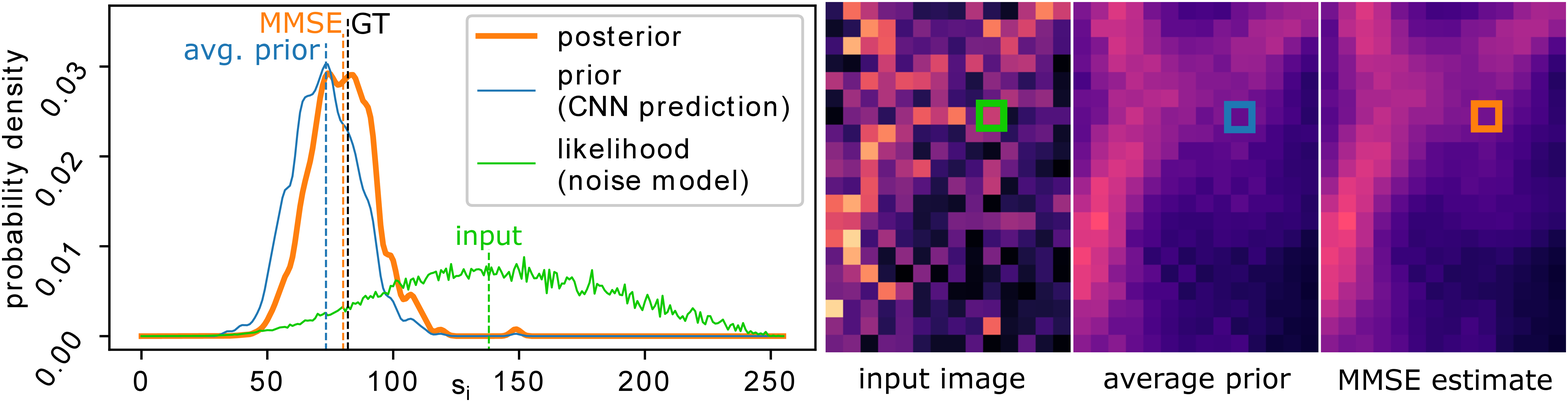}
\vspace{-10mm}
  \caption{
  Image denoising with PN2V.
  The final  MMSE estimate (orange dashed line) for the true signal $\sign_i$ of a pixel (position marked in the image insets on the right)
  corresponds to the center of mass of the posterior distribution (orange curve).
  Given an observed noisy input value $\img_i$ (dashed green line), the posterior
  is proportional to the product of the prior (blue curve) and the observation likelihood (green curve).
  PN2V describes the prior by a set of samples predicted by our CNN.
  The likelihood is provided by an arbitrary noise model.
  Black dashed line is the true signal of the pixel (GT).
  Prior and posterior are visualized using a kernel density estimator.
  }
  \label{fig:PixelDists}
  \vspace{-10mm}
\end{center}
\end{figure}
%
\section{Method}
\label{sec:method}
\compactsubsection{Maximum Likelihood Training}
\label{sec:training}
In PN2V, we build on the idea of masking pixels \cite{krull2019noise2voidCVPR} to obtain a prediction from the modified receptive field $\rfni$.
However, instead of directly predicting an estimate for each pixel value, PN2V trains a CNN to describe a probability distribution
\begin{equation}
    p(\sign_i|\rfni;\param). \label{eq:predictedProb}
\end{equation}
We will refer to $p(\sign_i|\rfni;\param)$ as \emph{prior}, as it describes our knowledge of the pixel's signal considering only its surroundings, but not the observation at the pixel itself $\img_i$, since it has been excluded from $\rfni$.
We choose a sample based representation for this prior, which will be discussed below.

Remembering that the observed pixels values are drawn independently (Eq.~\ref{eq:corruption}),
we can combine Eq.~\ref{eq:predictedProb} with our noise model, and obtain the joint distribution
\begin{equation}
    p(\img_i, \sign_i|\rfni;\param)=p(\sign_i|\rfni;\param) p(\img_i|\sign_i).
    \label{eq:joint}
\end{equation}
By integrating over all possible clean signals, we can derive
\begin{equation}
    \begin{split}
        p(\img_i|\rfni;\param) & = \int_{-\infty}^{\infty} p(\sign_i|\rfni;\param) p(\img_i|\sign_i) d{\sign_i},
    \end{split}
\end{equation}
the probability of observing the pixel value $\img_i$, given we know its surroundings $\rfni$.
We can now view CNN training as an unsupervised learning task.
Following the maximum likelihood approach, we tune $\param$ to minimize
\begin{equation}
    \begin{split}
        \argmin_{\param} \sum_{i=1}^n -\ln \left( \int_{-\infty}^{\infty} p(\sign_i|\rfni;\param) p(\img_i|\sign_i)  d{\sign_i} \right).
        \label{eq:optTask}
    \end{split}
\end{equation}
Note that in order to improve readability, we from here on omit the index $j$, and refrain from explicitly referring to the training image.

\compactsubsection{Sample Based Prior}
\label{sec:sample}
To allow an efficient optimization of Eq.~\ref{eq:optTask} we choose a sample based representation of our prior $p(\sign_i|\rfni;\param)$.
For every pixel $i$, our network directly predicts $K=800$ output values $\sign^k_i$, which we interpret as independent samples, drawn from $p(\sign_i|\rfni;\param)$.
We can now approximate Eq.~\ref{eq:optTask} as
\begin{equation}
    \begin{split}
        \argmin_{\param} \sum_{i=1}^n -\ln
        \left(
        \frac{1}{K}
        \sum_{k=1}^K p(\img_i|\sign^k_i)
        \right).
        \label{eq:sampleOptTask}
    \end{split}
\end{equation}
During training we use Eq.~\ref{eq:sampleOptTask} as loss function.
Note that the summation over $k$ can be efficiently performed on the GPU.

Since every sample $\sign^k_i$ is effectively a function of the parameters $\param$, we can calculate the derivative with respect to any network parameter $\param_l$ as
\begin{equation}
    \begin{split}
        \frac{\partial}{\partial \param_l}
        \sum_{i=1}^n -\ln
        \left(
        \frac{1}{K}
        \sum_{k=1}^K p(\img_i|\sign^k_i)
        \right) & =
        \sum_{i=1}^n
        -\left(
        \frac{\sum_{k=1}^K
        \frac{\partial}{\partial \sign^k_i} p(\img_i|\sign^k_i)
        \frac{\partial \sign^k_i}{\partial \param_l}
        }
        {\sum_{k=1}^K p(\img_i|\sign^k_i)}
        \right).
    \end{split}
\end{equation}
\compactsubsection{Minimal Mean Squared Error (MMSE) Inference}
\label{sec:inference}
Assuming our network is sufficiently trained, we are now interested in processing images and finding sensible estimates for every pixel's signal $\sign_i$.
Based on our probabilistic model, we derive the MMSE estimate, which is defined as
\begin{equation}
\begin{split}
    \signMMSE_i & = \argmin_{\hat{\sign}_i} \Esub{
    \left(\hat{\sign}_i -\sign_i \right)^2
    } {p(\sign_i|\rfi)}
     = \Esub{\sign_i } {p(\sign_i|\rfi)},
\end{split}
\end{equation}
where $p(\sign_i|\rfi)$ is the \emph{posterior} distribution of the signal given the complete surrounding patch.
The posterior is proportional to the joint distribution given in Eq.~\ref{eq:joint}.
We can thus approximate $\signMMSE_i$ by weighing our predicted samples with the corresponding observation likelihood and calculating their average
\begin{equation}
\begin{split}
    \signMMSE_i 
    & \approx
    \frac{
    \sum_{k=1}^K  p(\img_i|\sign^k_i) \sign^k_i
    } {
    \sum_{k=1}^K  p(\img_i|\sign^k_i)
    }.
\end{split}
\end{equation}
Figure~\ref{fig:PixelDists} illustrates the process and shows the involved distributions for a concrete pixel in a real example.
\section{Experiments}
\label{sec:experiments}
\begin{figure}[t]
\begin{center}
	  \includegraphics[width=1\textwidth]{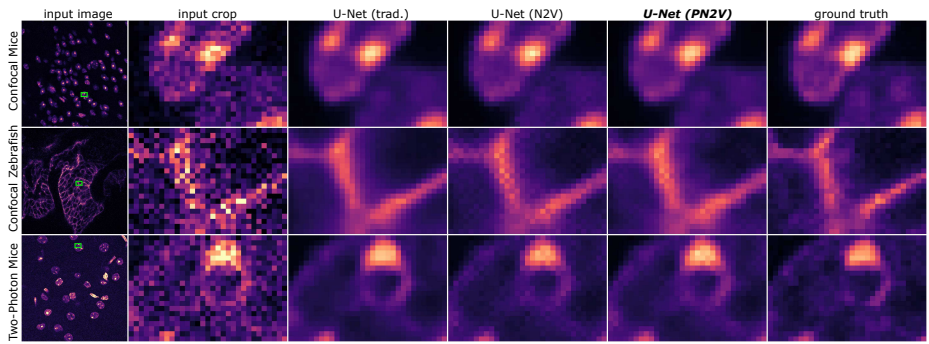}
\vspace{-7mm}
  \caption{Qualitative results for three images (rows) from the datasets we used in this manuscript.
  Left to right:~raw image (NR1), zoomed inset, predictions by U-Net (trad.), U-Net (N2V), U-Net (PN2V), and ground truth data. 
  }
  \label{fig:qualRes}
\end{center}
\end{figure}

The results of our experiments can be found in Table~\ref{tab:results}.
In Figure~\ref{fig:qualRes} we provide qualitative results on realistic test images.

\compactsubsection{Datasets}
\label{sub:datasets}
We evaluate PN2V on datasets provided by Zhang \etal in~\cite{zhang2018poisson}.
Since PN2V is not yet implemented for multi-channel images, we use all available single-channel datasets.

These datasets are recorded with different samples and under different imaging conditions.
Each of them consists of a total of 20 fields of view (FOVs).
One FOV is reserved for testing.
The other 19 are used for training and validation.

For each FOV, the data is composed of 50 raw microscopy images, each containing different noise realizations of the same static sample.
For every FOV, Zhang \etal additionally simulate four reduced noise regimes (NRs) by averaging different subsets of 2, 4, 8, and 16 raw images~\cite{zhang2018poisson}.
We will refer to the raw images as NR1 and to the regimes created through averaging 2, 4, 8, and 16 images as NR2, NR3, NR4, and NR5, respectively.

We find that in one of the datasets (\emph{Two-Photon Mice}) the average pixel intensity fluctuates heavily over the course of the 50 images, even though it should be approximately constant for each FOV.
Considering that a single ground truth image (the average) is used for the evaluation on all 50 images, this leads to fluctuations and distortions in the calculated PSNR values, which are also reflected in the comparatively high standard errors (SEMs) for all methods, see Table~\ref{tab:results}.
To account for this inconsistency in the data, we additionally use a variant of the PSNR calculation that is invariant to arbitrary shifts and linear transformations in the ground truth signal.
These values are marked by an asterisk (*).
Details can also be found in the supplementary material\footnote{
The Supplement will be made available soon.\vspace{-0.5cm}}.

\compactsubsection{Acquiring Noise Models}
\label{sec:noisemodel}
In our experiments, we use a histogram based method to measure and describe the noise distribution $p(\img_i|\sign_i)$.
We start with corresponding pairs of clean $\sign^j$ and noisy $\img^j$ images.
Here, we use the available training data from \cite{zhang2018poisson} for this purpose.
However, in general these images could show an arbitrary test pattern that covers the desired range of values and do not have to resemble the sample we are interested in.
We construct a 2D histogram (256$\times$256 bins), with the y- and x-axis corresponding to the clean $\sign^j_i$ and noisy pixel values $\img^j_i$, respectively.
By normalizing every row, we obtain a a probability distribution for every signal.
Considering Eq.~\ref{eq:sampleOptTask}, we require our model to be differentiable with respect to the $\sign_i$.
To ensure this differentiability, we linearly interpolate along the y-axis of the normalized histogram, obtaining a model for $p(\img_i|\sign_i)$ that is continuous in $\sign_i$.

\compactsubsection{Evaluated Denoising Models/Methods}
\label{sub:architecture}
To put the denoising results of PN2V into perspective, we compare to various state-of-the-art baselines, including the strongest published numbers on the datasets.

\compactsubsubsection{U-Net (PN2V)}
We use a standard U-Net~\cite{ronneberger2015u}.
Our network has a U-Net depth of 3, 1 input channel, and $K=800$ output channels, which are interpreted as samples.
We use a initial feature channel number of 64 in the fist U-Net layer.
We train our network separately for each NR in each dataset.
We use the same masking technique as \cite{krull2019noise2voidCVPR}.
Further details and training parameters can be found in the supplementary material\footnotemark[1].

\compactsubsubsection{U-Net (N2V)}
We use the same network architecture as for U-Net (PN2V) but modify the outputlayer to produce only a single prediction instead of $K=800$.
The network is trained using the Noise2Void scheme as described in \cite{krull2019noise2voidCVPR}.
All training parameters are identical to U-Net (PN2V).

\compactsubsubsection{U-Net (trad.)}
We use the exact same architecture as U-Net (N2V), but train the network using the available ground-truth data and the standard MSE loss (see Eq.~\ref{eq:risk}).
All training parameters are identical to U-Net (PN2V) and U-Net (N2V).

\compactsubsubsection{VST+BM3D}
Numbers are taken from \cite{zhang2018poisson}.
The authors fit a Poisson Gaussian noise model to the data and then apply a combination of variance-stabilizing transformation (VST)~\cite{makitalo2013optimal} and BM3D filtering~\cite{dabov2007image}.

\compactsubsubsection{DnCNN}
Numbers are taken from \cite{zhang2018poisson}.
DnCNN \cite{zhang2017beyond} is an established CNN based denoising architecture that is trained in a supervised fashion.

\compactsubsubsection{N2N}
Numbers are taken from \cite{zhang2018poisson}.
The authors train a network according to the N2N scheme, using an architecture similar to the one presented in \cite{noise2noise}.
%
%
\begin{table}[b]
\hspace{0.9mm}
\resizebox{.9\textwidth}{!}{\begin{minipage}{\textwidth}
\begin{tabular}{lllllll}
\multicolumn{7}{c}{\textBF{Confocal Mice}}                                                                                                                                                                                                                                                    \\ \cline{2-7} 
\multicolumn{1}{l|}{}               & \multicolumn{1}{c|}{NR1}                  & \multicolumn{1}{c|}{NR2}                  & \multicolumn{1}{c|}{NR3}                  & \multicolumn{1}{c|}{NR4}                  & \multicolumn{1}{c|}{NR5}                  & \multicolumn{1}{c|}{Mean}           \\ \hline
\multicolumn{1}{|l|}{Input}         & \multicolumn{1}{l|}{29.38$\pm$0.01}          & \multicolumn{1}{l|}{32.44$\pm$0.01}          & \multicolumn{1}{l|}{35.59$\pm$0.01}          & \multicolumn{1}{l|}{38.90$\pm$0.01}          & \multicolumn{1}{l|}{42.64$\pm$0.03}          & \multicolumn{1}{l|}{35.79}          \\ \hline
\multicolumn{1}{|l|}{U-Net (\textbf{\textit{PN2V}})}  & \multicolumn{1}{l|}{\textBF{38.24$\pm$0.02}} & \multicolumn{1}{l|}{\textBF{39.72$\pm$0.03}} & \multicolumn{1}{l|}{\textBF{41.34$\pm$0.03}} & \multicolumn{1}{l|}{\textBF{43.02$\pm$0.04}} & \multicolumn{1}{l|}{\textBF{45.11$\pm$0.05}} & \multicolumn{1}{l|}{\textBF{41.49}} \\
\multicolumn{1}{|l|}{U-Net (N2V)}   & \multicolumn{1}{l|}{37.56$\pm$0.02}          & \multicolumn{1}{l|}{38.78$\pm$0.02}          & \multicolumn{1}{l|}{39.94$\pm$0.02}          & \multicolumn{1}{l|}{41.01$\pm$0.02}          & \multicolumn{1}{l|}{41.95$\pm$0.02}          & \multicolumn{1}{l|}{39.85}          \\
\multicolumn{1}{|l|}{VST+BM3D}      & \multicolumn{1}{l|}{37.95}                & \multicolumn{1}{l|}{39.47}                & \multicolumn{1}{l|}{41.09}                & \multicolumn{1}{l|}{42.73}                & \multicolumn{1}{l|}{44.52}                & \multicolumn{1}{l|}{41.15}          \\ \hline
\multicolumn{1}{|l|}{U-Net (trad.)} & \multicolumn{1}{l|}{\textBF{38.38$\pm$0.02}} & \multicolumn{1}{l|}{\textBF{39.90$\pm$0.03}} & \multicolumn{1}{l|}{41.37$\pm$0.03}          & \multicolumn{1}{l|}{43.06$\pm$0.04}          & \multicolumn{1}{l|}{45.16$\pm$0.05}          & \multicolumn{1}{l|}{\textBF{41.58}} \\
\multicolumn{1}{|l|}{DnCNN}         & \multicolumn{1}{l|}{38.15}                & \multicolumn{1}{l|}{39.78}                & \multicolumn{1}{l|}{\textBF{41.41}}       & \multicolumn{1}{l|}{\textBF{43.11}}       & \multicolumn{1}{l|}{\textBF{45.20}}       & \multicolumn{1}{l|}{41.53}          \\
\multicolumn{1}{|l|}{N2N}           & \multicolumn{1}{l|}{38.19}                & \multicolumn{1}{l|}{39.77}                & \multicolumn{1}{l|}{41.28}                & \multicolumn{1}{l|}{42.83}                & \multicolumn{1}{l|}{44.56}                & \multicolumn{1}{l|}{41.33}          \\ \hline
\multicolumn{7}{c}{
\textBF{Confocal Zebrafish}}                                                                                                                                                                                                                                               \\ \hline
\multicolumn{1}{|l|}{Input}         & \multicolumn{1}{l|}{22.81$\pm$0.02}          & \multicolumn{1}{l|}{25.89$\pm$0.02}          & \multicolumn{1}{l|}{29.05$\pm$0.03}          & \multicolumn{1}{l|}{32.39$\pm$0.03}          & \multicolumn{1}{l|}{36.21$\pm$0.04}          & \multicolumn{1}{l|}{29.27}          \\ \hline
\multicolumn{1}{|l|}{U-Net (\textbf{\textit{PN2V}})}  & \multicolumn{1}{l|}{\textBF{32.45$\pm$0.02}} & \multicolumn{1}{l|}{\textBF{33.96$\pm$0.03}} & \multicolumn{1}{l|}{\textBF{35.48$\pm$0.05}} & \multicolumn{1}{l|}{\textBF{37.07$\pm$0.06}} & \multicolumn{1}{l|}{\textBF{39.08$\pm$0.07}} & \multicolumn{1}{l|}{\textBF{35.61}} \\
\multicolumn{1}{|l|}{U-Net (N2V)}   & \multicolumn{1}{l|}{32.10$\pm$0.02}          & \multicolumn{1}{l|}{33.34$\pm$0.03}          & \multicolumn{1}{l|}{34.43$\pm$0.04}          & \multicolumn{1}{l|}{35.39$\pm$0.04}          & \multicolumn{1}{l|}{36.21$\pm$0.03}          & \multicolumn{1}{l|}{34.30}          \\
\multicolumn{1}{|l|}{VST+BM3D}      & \multicolumn{1}{l|}{32.00}                & \multicolumn{1}{l|}{33.75}                & \multicolumn{1}{l|}{35.30}                & \multicolumn{1}{l|}{36.78}                & \multicolumn{1}{l|}{38.32}                & \multicolumn{1}{l|}{35.23}          \\ \hline
\multicolumn{1}{|l|}{U-Net (trad.)} & \multicolumn{1}{l|}{\textBF{32.93$\pm$0.03}} & \multicolumn{1}{l|}{34.35$\pm$0.04}          & \multicolumn{1}{l|}{35.67$\pm$0.05}          & \multicolumn{1}{l|}{37.11+0.06}           & \multicolumn{1}{l|}{\textBF{39.09$\pm$0.07}} & \multicolumn{1}{l|}{\textBF{35.83}} \\
\multicolumn{1}{|l|}{DnCNN}         & \multicolumn{1}{l|}{32.44}                & \multicolumn{1}{l|}{34.16}                & \multicolumn{1}{l|}{\textBF{35.75}}       & \multicolumn{1}{l|}{\textBF{37.28}}       & \multicolumn{1}{l|}{39.07}                & \multicolumn{1}{l|}{35.74}          \\
\multicolumn{1}{|l|}{N2N}           & \multicolumn{1}{l|}{\textBF{32.93}}       & \multicolumn{1}{l|}{\textBF{34.37}}       & \multicolumn{1}{l|}{35.71}                & \multicolumn{1}{l|}{37.06}                & \multicolumn{1}{l|}{38.65}                & \multicolumn{1}{l|}{35.74}          \\ \hline
\multicolumn{7}{c}{\textBF{Two-Photon Mice}}                                                                                                                                                                                                                                                  \\ \hline
\multicolumn{1}{|l|}{Input}         & \multicolumn{1}{l|}{24.94$\pm$0.07}          & \multicolumn{1}{l|}{27.83$\pm$0.1}           & \multicolumn{1}{l|}{30.69$\pm$0.15}          & \multicolumn{1}{l|}{33.67$\pm$0.19}          & \multicolumn{1}{l|}{37.72$\pm$0.14}          & \multicolumn{1}{l|}{30.97}          \\ \hline
\multicolumn{1}{|l|}{U-Net (\textbf{\textit{PN2V}})}  & \multicolumn{1}{l|}{33.67$\pm$0.33}          & \multicolumn{1}{l|}{34.58$\pm$0.39}          & \multicolumn{1}{l|}{35.42$\pm$0.42}          & \multicolumn{1}{l|}{36.58$\pm$0.37}          & \multicolumn{1}{l|}{\textBF{39.78$\pm$0.24}} & \multicolumn{1}{l|}{36.00}          \\
\multicolumn{1}{|l|}{U-Net (N2V)}   & \multicolumn{1}{l|}{33.42$\pm$0.31}          & \multicolumn{1}{l|}{34.31$\pm$0.36}          & \multicolumn{1}{l|}{35.09$\pm$0.38}          & \multicolumn{1}{l|}{36.08$\pm$0.33}          & \multicolumn{1}{l|}{37.80$\pm$0.14}          & \multicolumn{1}{l|}{35.34}          \\
\multicolumn{1}{|l|}{VST+BM3D}      & \multicolumn{1}{l|}{\textBF{33.81}}       & \multicolumn{1}{l|}{\textBF{34.78}}       & \multicolumn{1}{l|}{\textBF{35.77}}       & \multicolumn{1}{l|}{\textBF{36.97}}       & \multicolumn{1}{l|}{39.39}                & \multicolumn{1}{l|}{\textBF{36.14}} \\ \hline
\multicolumn{1}{|l|}{U-Net (trad.)} & \multicolumn{1}{l|}{\textBF{34.35$\pm$0.19}}        & \multicolumn{1}{l|}{35.32+0.23}           & \multicolumn{1}{l|}{36.14$\pm$0.27}          & \multicolumn{1}{l|}{\textBF{37.48$\pm$0.27}}          & \multicolumn{1}{l|}{40.28$\pm$0.2}           & \multicolumn{1}{l|}{\textBF{36.72}}          \\
\multicolumn{1}{|l|}{DnCNN}         & \multicolumn{1}{l|}{33.67}                & \multicolumn{1}{l|}{34.95}                & \multicolumn{1}{l|}{36.10}                & \multicolumn{1}{l|}{37.43}                & \multicolumn{1}{l|}{\textBF{40.30}}       & \multicolumn{1}{l|}{36.49}          \\
\multicolumn{1}{|l|}{N2N}           & \multicolumn{1}{l|}{34.33}                & \multicolumn{1}{l|}{\textBF{35.32}}       & \multicolumn{1}{l|}{\textBF{36.25}}       & \multicolumn{1}{l|}{37.46}                & \multicolumn{1}{l|}{39.89}                & \multicolumn{1}{l|}{36.65}          \\ \hline
\hline
\multicolumn{1}{|l|}{$*$ U-Net (\textbf{\textit{PN2V}})}  & \multicolumn{1}{l|}{\textBF{34.84$\pm$0.06}} & \multicolumn{1}{l|}{\textBF{36.02$\pm$0.07}} & \multicolumn{1}{l|}{\textBF{37.08$\pm$0.08}} & \multicolumn{1}{l|}{\textBF{38.28$\pm$0.09}} & \multicolumn{1}{l|}{\textBF{40.89$\pm$0.07}} & \multicolumn{1}{l|}{\textBF{37.42}} \\
\multicolumn{1}{|l|}{$*$ U-Net (N2V)}   & \multicolumn{1}{l|}{34.60$\pm$0.09}          & \multicolumn{1}{l|}{35.77$\pm$0.1}           & \multicolumn{1}{l|}{36.71$\pm$0.1}           & \multicolumn{1}{l|}{37.64$\pm$0.09}          & \multicolumn{1}{l|}{38.49$\pm$0.05}          & \multicolumn{1}{l|}{36.64}          \\ \hline
\multicolumn{1}{|l|}{$*$ U-Net (trad.)} & \multicolumn{1}{l|}{35.05$\pm$0.05}          & \multicolumn{1}{l|}{36.22$\pm$0.06}          & \multicolumn{1}{l|}{37.28$\pm$0.07}          & \multicolumn{1}{l|}{38.78$\pm$0.1}           & \multicolumn{1}{l|}{41.34$\pm$0.07}          & \multicolumn{1}{l|}{37.73}          \\ \hline
\end{tabular}
\end{minipage}}
\\
\caption{
Results of PN2V and baseline methods on three datasets from~\cite{zhang2018poisson}.
Comparisons are performed on five noise regimes (NR1-NR5).
Numbers report PSNR (dB) $\pm$ 2 SEM, averaged over all $50$ images in each NR.
We group all supervised/non-supervised methods and mark the highest values in bold.
Rows marked by asterisk ($*$) use a scale- and shift-invariant PSNR calculation to address inconsistent acquisitions in the \emph{Two-Photon mice} dataset (see main text).
\textit{\textbf{Comp. times:}} All CNN based methods required below 1s per image (\emph{NVIDIA TITAN Xp}); VST+BM3D required on avg. 6.22s.
\label{tab:results}
}
\vspace{-.5cm}
\end{table}
\vspace{-5cm}
%
\section{Discussion}
\label{sec:discussion}
We have introduced PN2V, a fully probabilistic approach extending self-supervised CARE training.
PN2V makes use of an arbitrary noise model which can be determined by analyzing any set of available images that are subject to the same type of noise.
This is a decisive advantage compared to state-of-the-art supervised methods and allows PN2V to be used for many practical applications.

The much improved performance of PN2V lies consistently beyond self-supervised training and can often compete with state-of-the-art supervised methods.
We see a plethora of unique applications for PN2V, for example in challenging low-light conditions, where noise typically is the limiting factor for downstream analysis.

\subsection*{Acknowledgements}
We thank Uwe Schmidt, Martin Weigert, and Tobias Pietzsch for the helpful discussions.
We thank Tobias Pietzsch for proof reading.
The computations were performed
on an HPC Cluster at the Center for Information Services
and High Performance Computing (ZIH) at TU Dresden.
%
{\small
\enlargethispage{2\baselineskip}
\bibliographystyle{splncs04}
\bibliography{bib}
}
\end{document}